\documentclass[
    ,final            
  ]
  {aipproc}

\layoutstyle{8x11single}
\usepackage{graphicx}

\begin{document}

\title{Ultra-compact X-ray binaries in the Galactic Bulge}


\author{Lennart~van~Haaften}{
  address={Department of Astrophysics/ IMAPP, Radboud University Nijmegen, P.O. Box 9010, 6500 GL Nijmegen, The Netherlands}
}
\author{Gijs~Nelemans}{
}

\begin{abstract}
We model the present-day number and characteristics of Ultra-compact X-ray binaries (UCXBs) in the Galactic Bulge. The main objective is to compare the results with the forthcoming data from the Galactic Bulge Survey (GBS) in order to learn about formation scenarios, notably core collapse supernova rate, common envelope evolution and dynamical interactions, as well as the binary initial mass function. We use a binary population synthesis code and detailed stellar evolutionary tracks, combined with observations of the 11 known UCXBs. We predict a few hundred faint UCXBs with periods of $20-40\ \mbox{minutes}$ that would be detectable by the GBS.
\end{abstract}

\maketitle

\section{Introduction}

Ultra-compact X-ray binaries (UCXBs) are low mass X-ray binaries with an orbital period below ca. $90\ \mbox{min}$, indicating a compact and therefore hydrogen deficient donor \cite{nelson1986}. Via Roche Lobe overflow (RLOF) the donor loses mass, which is partially accreted by a very compact companion (a neutron star or a black hole). Due to friction in the accretion disk, the gas heats up to temperatures corresponding to X-ray radiation.

The Galactic Bulge is the best direction to look in to find UCXBs because of the high local star concentration and the fact that the line of sight from Earth crosses the Galactic Disk. Also, UCXBs are too rare for us to find nearby in the Milky Way, and too faint to resolve in other galaxies except for the closest.

\begin{figure}[h]
  \includegraphics[height=60mm]{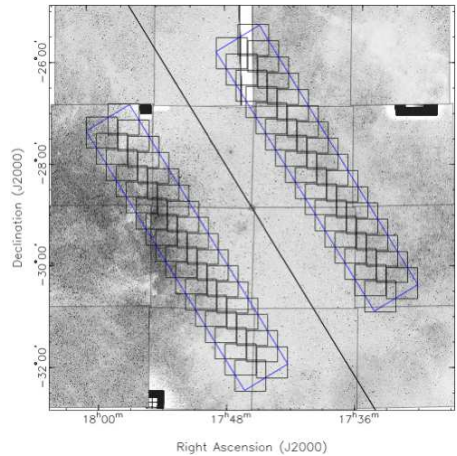}
  \caption{The target area of the Galactic Bulge Survey consists of two strips of $6^{\circ} \times 1^{\circ}$ around the Galactic Center. Image courtesy of P.~Jonker.}
  \label{fig:gbs}
\end{figure}

The Galactic Bulge Survey \cite{jonker2009} (GBS) is an X-ray and optical survey focused on a $12$ square degree region around the center of the Milky Way (fig. \ref{fig:gbs}). The aim of this study is to predict and explain observations made by the GBS, and gain a better understanding of the physical processes in UCXBs.

\section{Method}

The binary population synthesis code SeBa \cite{nelemans2001} has been used to evolve a population of zero-age main sequence binaries. From the evolved population two classes of systems are selected, with donors being white dwarfs or helium burning stars. Systems must have a neutron star or black hole as accretor. A third formation scenario involves main sequence donors and magnetic capture \cite{sluys2005} but this class is excluded here due to the comparatively low number of ultra-compact systems that appear in this way.

For the subsequent evolution more detailed stellar evolutionary tracks are used. For the white dwarf donors, a zero-temperature approximation \cite{marsh2004} has been used after the star has cooled sufficiently. The tracks for helium burning donor are taken from \cite{yungelson2008}.

The star formation history of the Galactic Bulge is subsequently used to arrive at the present-day number and characteristics of UCXBs in the Bulge. Using longterm observations by the Rossi XTE All Sky Monitor (ASM), these model parameters can be converted to observational parameters, which corresponds to a prediction of the present-day observable population.

\section{Results}

\subsection{Population synthesis}

Helium burning donor systems are typically formed less than $500\ \mbox{Myr}$ after system formation, and start RLOF within another $200\ \mbox{Myr}$, while the majority of white dwarf donor systems start RLOF only after $1-6\ \mbox{Gyr}$. Part of the explanation is that helium burning stars evolve from secondaries with a zero-age main sequence mass of $3-4\ M_{\odot}$ and therefore have a shorter main sequence timescale than the least massive white dwarf progenitors have. Also, if the orbit at the time of the formation of the helium burning star is too wide, it will turn into a white dwarf before starting RLOF, disqualifying itself from the sample and leaving only systems take reach RLOF relatively early. The primary explodes in a supernova early on, after typically a few $\times\ 10\ \mbox{Myr}$. Most of the white dwarf donor systems reach RLOF within $2\ \mbox{Gyr}$, but a significant fraction of the systems take much longer. This is the case for both helium and carbon-oxygen white dwarfs. Helium white dwarf donors typically arise from approximately $2\ M_{\odot}$ secondaries. Carbon-oxygen white dwarf donors mainly evolve from $3-10\ M_{\odot}$ primaries or secondaries that do undergo a helium flash. Unlike helium burning stars, white dwarfs can take arbitrarily long to reach RLOF.

\begin{figure}
  \includegraphics[height=60mm]{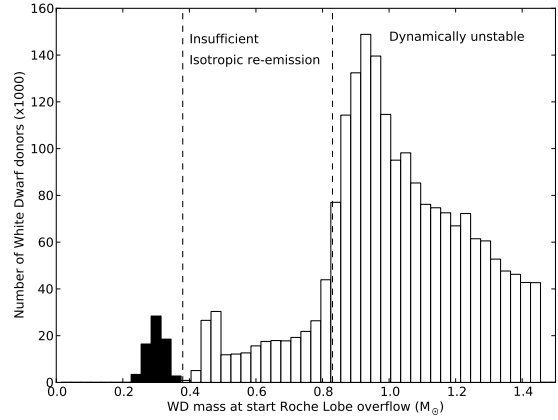}
  \caption{Total time-integrated amount of UCXBs with a white dwarf donor. Black: helium white dwarf donors. White: carbon-oxygen white dwarf donors.}
  \label{fig:wd}
\end{figure}

Many white dwarfs will merge upon RLOF (fig. \ref{fig:wd}). White dwarfs with a neutron star companion and a mass above $0.83\ M_{\odot}$ experience dynamically unstable RLOF, leading to runaway mass loss and a common envelope and eventually a merger. For a black hole accretor, this limit is approximately $1.0\ M_{\odot}$. Of the remaining systems, the ones with a mass over $0.38\ M_{\odot}$ cannot eject enough of the transfered matter from the system by isotropic re-emission. Even though the system is stable from the donor perspective, it is not from the accretor perspective and the system will still end up merging. In effect, almost all carbon-oxygen white dwarf systems merge, while almost all helium white dwarf systems do not merge. The exception are systems with black hole accretors. In that case low mass carbon-oxygen white dwarf systems can experience stable mass transfer. Systems that are unstable in either way are removed from the sample.

\subsection{Star formation history and birthrates}

\begin{figure}
  \includegraphics[height=60mm]{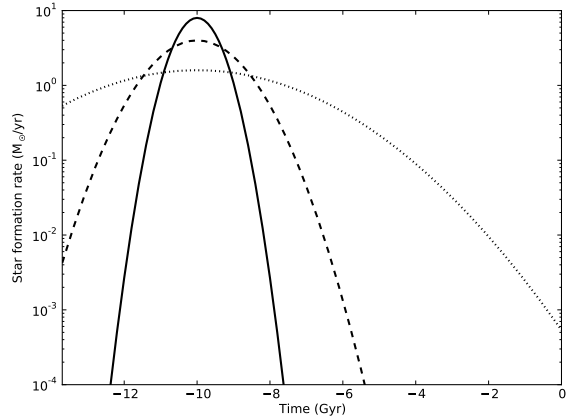}
  \caption{Star formation history of the Galactic Bulge as a Gaussian distribution for $\mu = -10\ \mbox{Gyr}$ and $\sigma = 0.5\ \mbox{(solid)}, 1.0\ \mbox{(dashed) and}\ 2.5\ \mbox{(dotted)\ Gyr}$. The total star formation is $1 \cdot 10^{10}\ M_{\odot}$. Note that $\mbox{Time} = 0$ corresponds to the present. \cite{clarkson2009,wyse2009}}
  \label{fig:sfh}
\end{figure}

Besides knowing which systems form and how long it takes for them to reach RLOF, we also need to know when these systems form: this is determined by the star formation history of the Bulge, shown in fig. \ref{fig:sfh}. The star formation is mostly concentrated around $10\ \mbox{Gyr}$ ago, but important is the tail of recent star formation in the $\sigma=2.5\ \mbox{Gyr}$ distribution.

\begin{figure}
  \centerline{
    \mbox{\includegraphics[height=55mm]{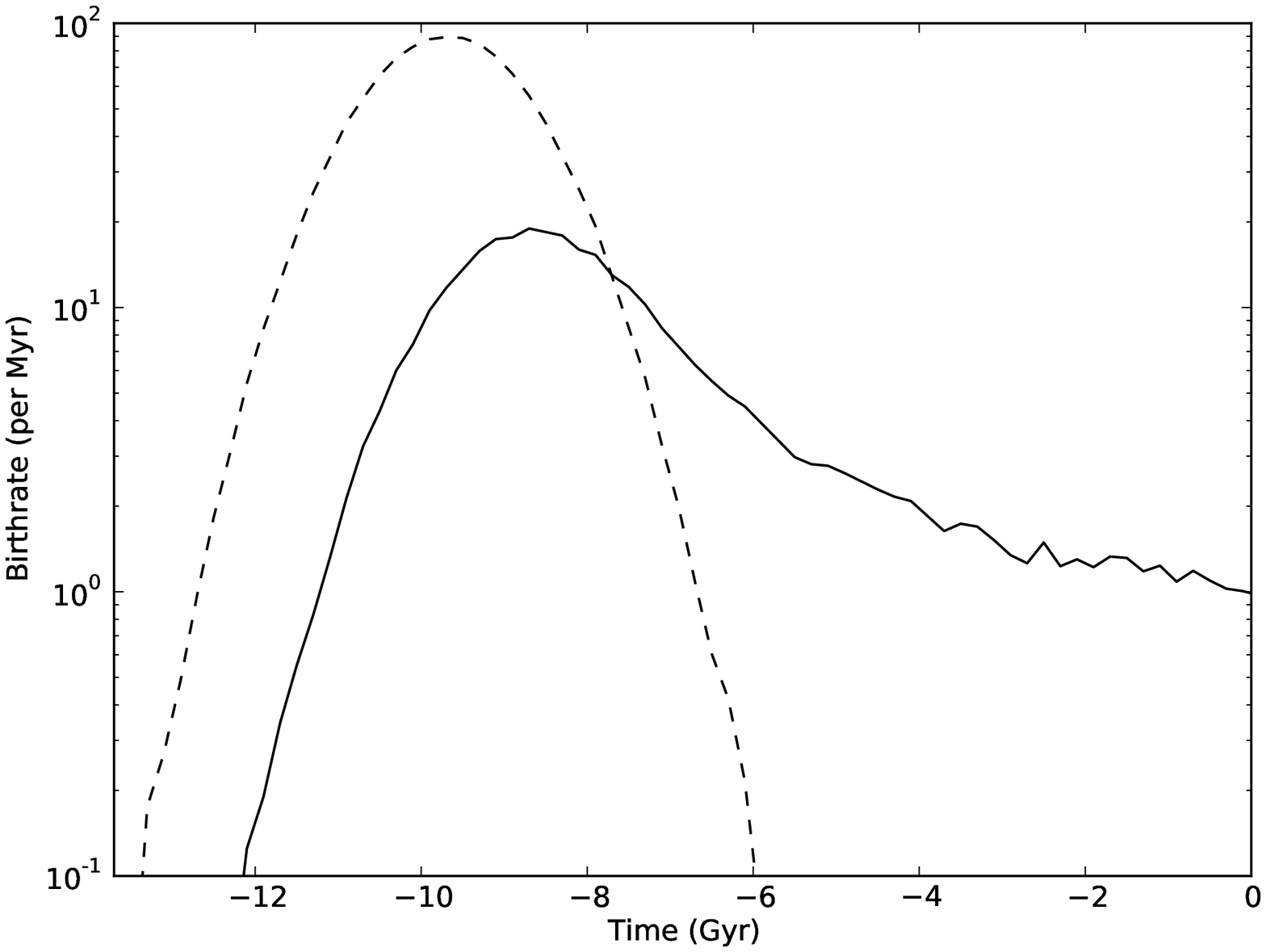}}
    \mbox{\includegraphics[height=55mm]{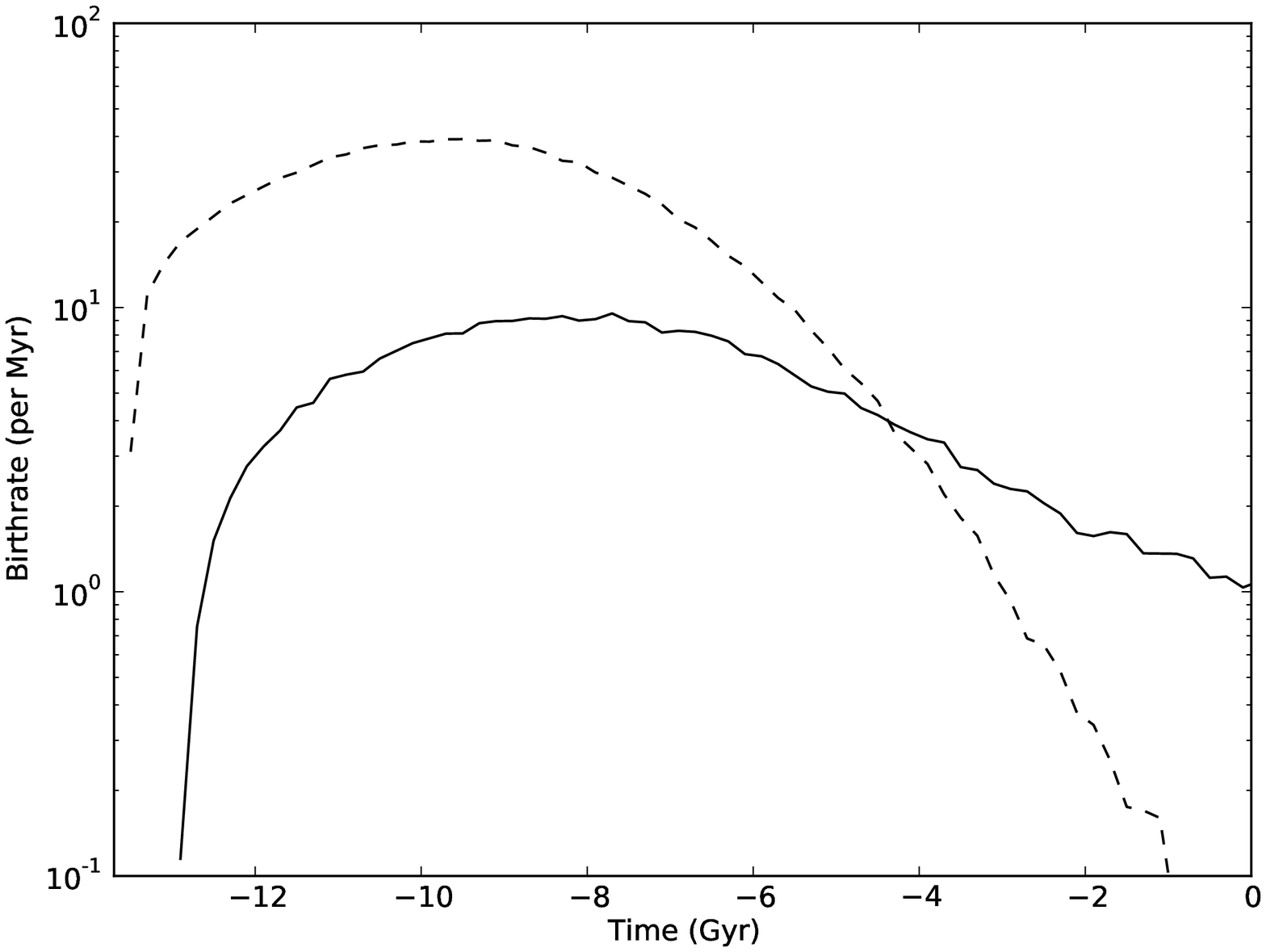}}
  }
  \caption{Birthrates for white dwarf donor systems (solid) and helium burning donor systems (dashed). Left: $\sigma = 1\ \mbox{Gyr}$. Right: $\sigma = 2.5\ \mbox{Gyr}$.}
  \label{fig:brate}
  \end{figure}

Combining the star formation history with the delay times of the start of RLOF yields birthrates for all systems, as shown in fig. \ref{fig:brate}. The main difference is that white dwarfs can start RLOF recently, while helium burning donor systems cannot, unless there is significant recent star formation ($\sigma=2.5\ \mbox{Gyr}$).

The evolution following RLOF or shortly before has been computed using tracks based on stellar evolution models such as \cite{yungelson2008}. For this specific stage of evolution, these tracks are more detailed than SeBa and therefore preferred. The evolution of UCXBs with white dwarf donors is governed by angular momentum loss by gravitational wave radiation. Mass loss from the donor is required to keep the radii of the donor and its Roche Lobe in accordance. After RLOF the donor radius will increase immediately for (fully degenerate) white dwarf donors and after roughly $100\ \mbox{Myr}$ for helium burning donors (once the donor has become sufficiently degenerate following the extinction of nuclear fusion due to mass loss). The orbital period will increase with mass loss for degenerate donors.

\begin{figure}
  \includegraphics[height=60mm]{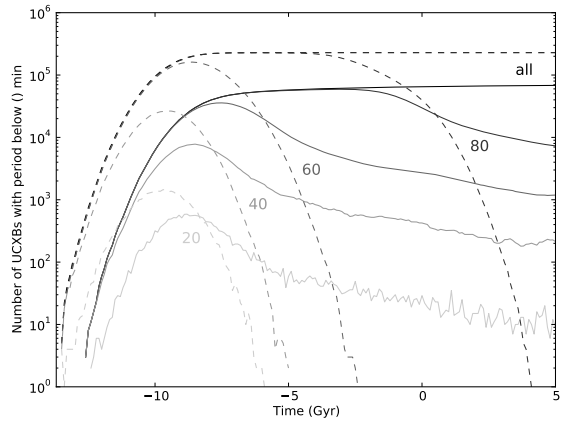}
  \caption{Amount of UCXBs against time. Solid: white dwarf donors. Dashed: helium burning donors. $\sigma = 1\ \mbox{Gyr}$.}
  \label{fig:history}
\end{figure}

The amount of systems below a given period (fig. \ref{fig:history}) initially follows the star formation rate as more systems are formed at short periods. As the star formation slows down, the amount keeps going up for a short while because more new systems are formed than older systems are removed from the given sample due to increasing periods. Given enough time, the amount will go down. Notably, the amounts of helium burning donor systems and white dwarf donor systems decline at different rates corresponding to their recent birthrates (fig. \ref{fig:brate}).

\subsection{Present Bulge population}

\begin{figure}
  \centerline{
    \mbox{\includegraphics[height=55mm]{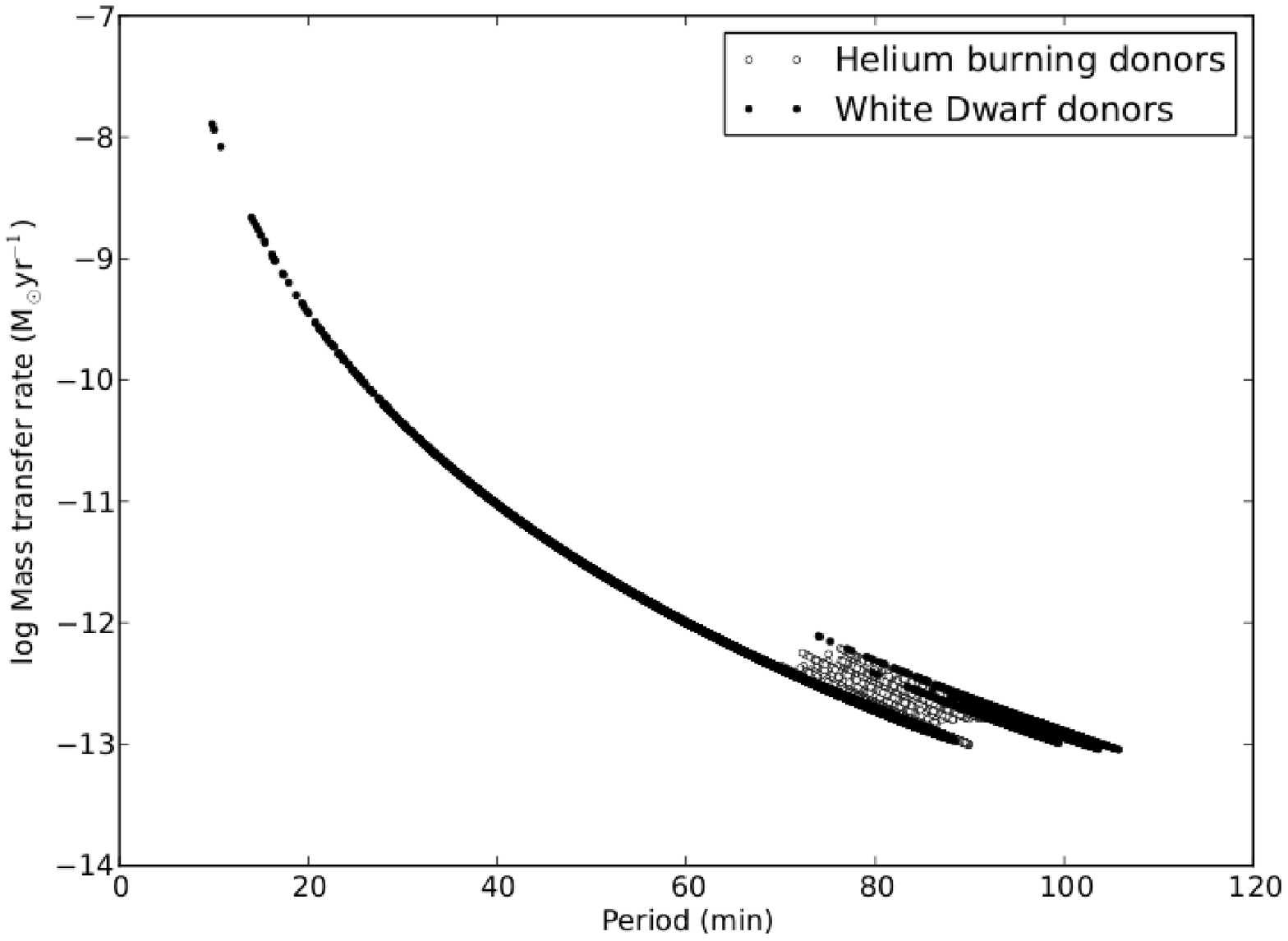}}
    \mbox{\includegraphics[height=55mm]{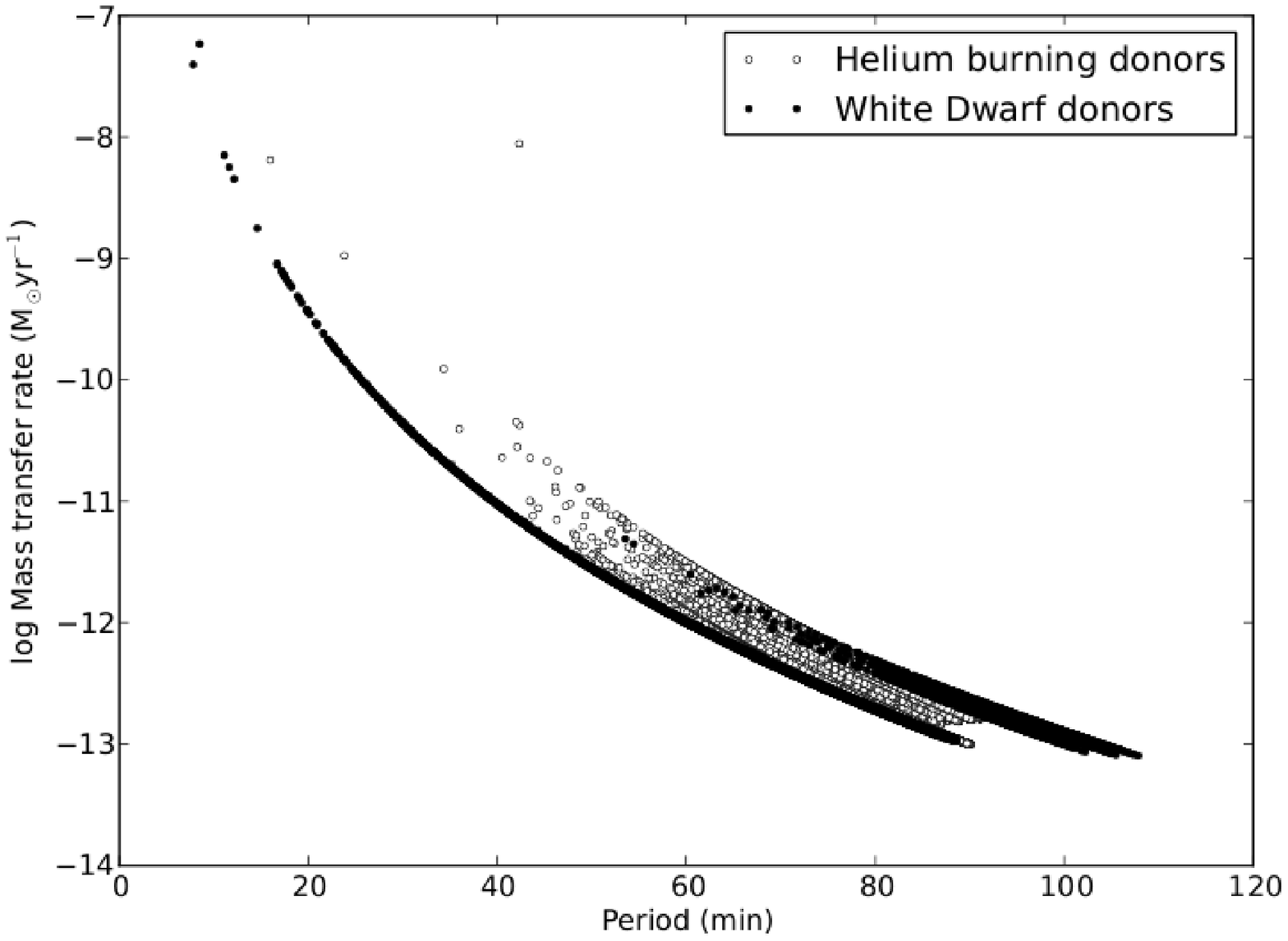}}
  }
  \caption{Present-day orbital periods and mass transfer rates for white dwarf donors (black) and helium burning donors (white). Left: $\sigma = 1\ \mbox{Gyr}$. Right: $\sigma = 2.5\ \mbox{Gyr}$.}
  \label{fig:permdot}
  \end{figure}

Of special interest is the population today. For a star formation history distribution $\sigma = 1\ \mbox{Gyr}$, most systems are old and have expanded to a period of around $80\ \mbox{min}$ (fig. \ref{fig:permdot}). All short period systems are young and therefore must contain white dwarf donors, generally helium white dwarfs. White dwarfs with black hole companions reach periods longer than $90\ \mbox{min}$ because the gravitational wave timescale is shorter for higher accretor mass, resulting in a higher mass transfer rate and a lower donor mass than for a less massive accretor, after a given time. For $\sigma=2.5\ \mbox{Gyr}$ young helium burning donor systems exist as well, because of the recent star formation. Their donors have turned into carbon-oxygen white dwarfs during their evolution.

\begin{figure}
  \includegraphics[height=60mm]{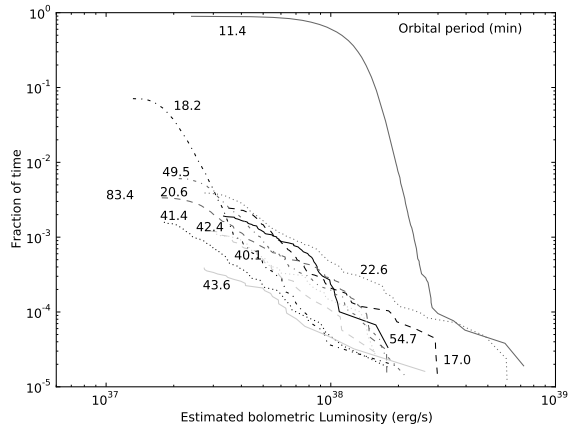}
  \caption{UCXB variability: fraction of time that a source emits at a given luminosity for 11 known UCXBs. \cite{liu2008,krimm2007,zand2007}}
  \label{fig:signi}
\end{figure}

We are interested in what we can observe and therefore must convert model parameters to observables. More specifically, we have to convert modeled mass transfer rates to observational luminosities. A reliable method is to use observations of known UCXBs by the ASM. The ASM has collected count rates in the $2-10\ \mbox{keV}$ range since $1996$. These have been converted to bolometric luminosities using distance estimates from \cite{liu2008,krimm2007} and a photon energy and bolometric conversion by \cite{zand2007}. Figure \ref{fig:signi} shows how often a source emits above a given luminosity. Most UCXBs are highly variable. Most systems are only visible a small fraction of the time above a given luminosity, so at any given time we will only see a small fraction of the UCXB population using an instrument with a given sensitivity. The short period systems are generally visible more often than the long period systems, which is to be expected given their higher mass transfer rates.

\begin{figure}
  \includegraphics[height=60mm]{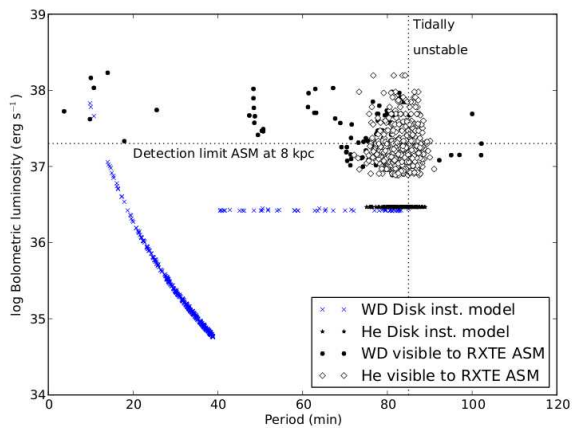}
  \caption{Predicted observable population of UCXBs, based on ASM observations and DIM theory. $\sigma = 1\ \mbox{Gyr}$.}
  \label{fig:perlum}
\end{figure}

This data can be applied to the modeled systems. Depending on their period, a luminosity is assigned, drawn randomly from the time-luminosity curve (fig. \ref{fig:signi}) of the best-matching real UCXB. More than $99 \%$ of the modeled systems will be invisible at any given time. This causes a "population inversion": even though there really exist many more long period (around $80\ \mbox{min}$) UCXBs, from a observational point of view there are more short period systems.
Applying the ASM data on the model results (fig. \ref{fig:perlum}, $\sigma = 1\ \mbox{Gyr}$) gives the circles above the ASM sensitivity limit. The data points represent the prediction of the present-day population of UCXBs. Some circles are below this limit because some real UCXBs are closer to us than the Galactic Center. In general though, below the ASM limit no prediction is possible using the ASM data. In a forthcoming paper these observational results will be explained in greater detail.
A second method to convert mass transfer rate to luminosity is the Disk Instability Model (DIM) \cite{zand2007}, which can be used to calculate luminosities fainter than ASM can observe. According to this theory, a mass transfer rate must exceed a critical value in order to be stable and the source to be persistent, i.e. visible at a high luminosity (almost) all the time. For lower mass transfer rates, the accretion disk builds until the cold outer disk becomes unstable and collapses onto the accretor. The source is only visible during these outburst stages, therefore the fraction of time the source is visible goes down for lower mass transfer rate, while the luminosity during these outbursts is held at a fixed value.

At long periods, the actual number of systems will be much less than shown in fig. \ref{fig:perlum} due to tidal disruption. Tidal heating will change the mass-radius relation of the low mass donor leading to disruption of the system \cite{ruderman1985}. For $\sigma = 2.5\ \mbox{Gyr}$, there are more carbon-oxygen donors at $40-60\ \mbox{min}$ which have evolved from helium burning donors.

\section{Discussion and Conclusions}

\begin{itemize}
\item Tidal instability for low mass donors may be more effective than theory suggests, otherwise the predicted number of bright long period UCXBs is much larger than the observed.
\item Since the observed UCXBs (periods below $60\ \mbox{min}$) with known composition consist of helium and carbon-oxygen in more or less equal numbers \cite{nelemans2010}, recent star formation is required which contrains the star formation history model. A carbon-oxygen donor implies a helium burning donor origin because carbon-oxygen white dwarf donor systems merge upon RLOF in case of a neutron star accretor, and a black hole accretor system would likely be disrupted already unless it is very young. Short period (below $20\ \mbox{min}$) UCXBs are most likely helium white dwarfs because they have a higher formation rate at the present, regardless of which star formation history is used.
\item Prediction: the existence of a few hundred faint UCXBs with periods between $20-40\ \mbox{min}$ that would be observable by the GBS.
\end{itemize}

\begin{theacknowledgments}
We would like to thank Rasmus Voss, Lev Yungelson and Marc van der Sluys for their help and contributions. This study was supported by the Netherlands Organisation for Scientific Research (NWO).
\end{theacknowledgments}

\bibliographystyle{aipproc}
\bibliography{lennart_refs}

\end{document}